\documentclass[12pt]{article}
\usepackage{amsmath}
\usepackage{times}
\usepackage{graphicx}
\usepackage{color}
\usepackage{multirow}
\usepackage[authoryear]{natbib}
\usepackage{rotating}
\usepackage{bbm}
\usepackage{latexsym}
\usepackage{bm}
\usepackage{amsfonts}
\usepackage{subcaption}
\usepackage[export]{adjustbox}
\usepackage{subfiles}

\usepackage[ruled,vlined]{algorithm2e}

\newcommand{\captionfonts}{\normalsize}

\makeatletter  
\long\def\@makecaption#1#2{%
  \vskip\abovecaptionskip
  \sbox\@tempboxa{{\captionfonts #1: #2}}%
  \ifdim \wd\@tempboxa >\hsize
    {\captionfonts #1: #2\par}
  \else
    \hbox to\hsize{\hfil\box\@tempboxa\hfil}%
  \fi
  \vskip\belowcaptionskip}
\makeatother   

\usepackage[skip=0pt]{caption} 
\usepackage[symbol]{footmisc}

\usepackage{setspace}
\begin{document}
\hspace{13.9cm}1

\ \vspace{20mm}\\

\singlespacing

{\large \bf \centering ReBaCCA-ss: Relevance-Balanced Continuum Correlation Analysis with Smoothing and Surrogating for Quantifying Similarity Between Population Spiking Activities}

\ \\
{\large \bf{Xiang Zhang\footnotemark[1]{}} \\
\it xzhang21@usc.edu \\
Department of Biomedical Engineering, University of Southern California, Los Angeles, California, United States
\\} \\
{\large \bf{Chenlin Xu} \\
\it jasonchenlinxu@outlook.com \\
Department of Biomedical Engineering, Viterbi School of Engineering, University of Southern California, Los Angeles, California, United States
\\} \\
{\large \bf{Zhouxiao Lu} \\
\it luzhouxi@usc.edu \\
Department of Biomedical Engineering, Viterbi School of Engineering, University of Southern California, Los Angeles, California, United States
\\} \\
{\large \bf{Haonan Wang} \\
\it wanghn@stat.colostate.edu \\
Department of Statistics, Colorado State University, Fort Collins, Colorado State, United States
\\} \\
{\large \bf{Dong Song\footnotemark[1]{}} \\
\it dsong@usc.edu \\
Department of Neurological Surgery, Keck School of Medicine; Department of Biomedical Engineering, Viterbi School of Engineering, University of Southern California, Los Angeles, California, United States
} 

\footnotetext[1]{Xiang Zhang and Dong Song are the corresponding authors} 

\thispagestyle{empty}
\markboth{}{NC instructions}
\ \vspace{-0mm}\\

\newpage
\begin{center} {\bf Abstract} \end{center}
{\bf
Quantifying similarity between population spike patterns is essential for understanding how neural dynamics encode information. Traditional approaches, which combine kernel smoothing, PCA, and CCA, have limitations: smoothing kernel bandwidths are often empirically chosen, CCA maximizes alignment between patterns without considering the variance explained within patterns, and baseline correlations from stochastic spiking are rarely corrected. We introduce ReBaCCA-ss (Relevance-Balanced Continuum Correlation Analysis with smoothing and surrogating), a novel framework that addresses these challenges through three innovations: (1) balancing alignment and variance explanation via continuum canonical correlation; (2) correcting for noise using surrogate spike trains; and (3) selecting the optimal kernel bandwidth by maximizing the difference between true and surrogate correlations. ReBaCCA-ss is validated on both simulated data and hippocampal recordings from rats performing a Delayed Nonmatch-to-Sample task. It reliably identifies spatio-temporal similarities between spike patterns. Combined with Multidimensional Scaling, ReBaCCA-ss reveals structured neural representations across trials, events, sessions, and animals, offering a powerful tool for neural population analysis.
}


\doublespacing
\newpage
\section{Introduction}
 The brain encodes and transmits information through population spiking activities --- spatial-temporal patterns of neuronal spikes \citep{pillow2017population,saxena2019towards,zhang2019clustering,she2022double,she2024distributed}. Spatial patterns reflect coordinated interactions across neurons, while temporal patterns capture how neurons respond to stimuli over time. Together, these dynamics provide a foundation for understanding the brain’s information processing. Recent advances in electrode technology enable large-scale simultaneous recordings of hundreds to thousands of neurons, significantly increasing the availability of population spiking data and opening new possibilities to understand brain functions \citep{jonsson2016bioelectronic,xu2018acute,scholten2023shared,chen2023recent}.
 
A crucial aspect of analyzing population spiking activities is quantifying the similarity between two spatial-temporal patterns. For example, comparing spiking data from the same subject across different tasks can reveal how neural populations adapt to new conditions. Cross-subject comparisons can help determine whether a universal computational strategy governs neural spiking activity. However, measuring such similarity presents significant technical challenges. 

Spatially, there is no direct neuron-to-neuron correspondence between subjects \citep{williamson2019bridging}, and even within the same subject, neurons may appear or disappear. The high dimensionality of neuronal populations further complicates comparisons, as spike trains often contain redundant information arising from low-dimensional latent dynamics \citep{gallego2020long,safaie2023preserved}, making direct comparisons less meaningful.

Temporally, spike trains are sparse point processing signals \citep{truccolo2005point,chen2019stability,huang2022extracting}. Direct correlation calculations typically yield near-zero values, making them ineffective for meaningful comparisons. Furthermore, the stochastic nature of neural firing can produce spikes by chance, leading to spurious patterns that could be misinterpreted as meaningful activity.

A common approach in neuroscience to address these challenges is to integrate kernel smoothing, dimensionality reduction, and cross-dataset alignment \citep{gallego2020long,safaie2023preserved}. First, point-process spike trains are converted into continuous signals (firing rates) by convolving them with a smoothing kernel of a specified bandwidth. Next, Principal Component Analysis (PCA) is applied to extract low-dimensional latent dynamics from these signals. Finally, Canonical Correlation Analysis (CCA) is used to align latent dynamics across different recording sessions or subjects. The similarity of the spike pattern is then quantified as the mean of the top Canonical Correlations (CCs). This approach has provided valuable insights into how low-dimensional latent dynamics shape behavioral execution over time \citep{gallego2020long} and across animals \citep{safaie2023preserved}.

\begin{figure}[!htbp]
\begin{center}
\includegraphics[width=\textwidth]{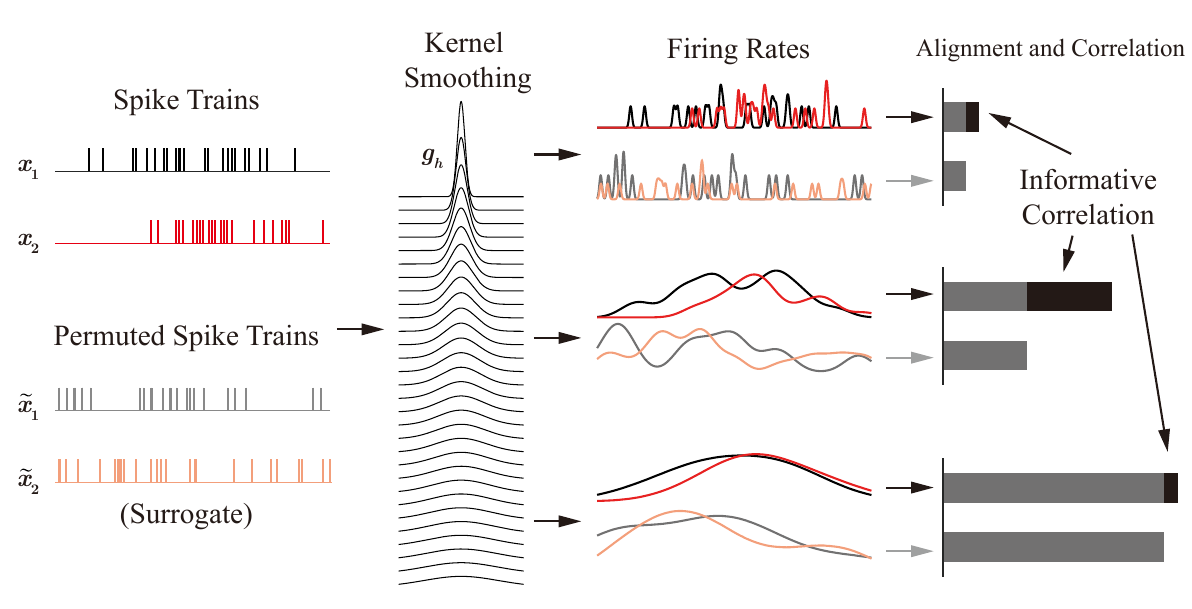}
\end{center}
\caption{Quantifying similarity between spike trains with kernel smoothing. Two spike trains, $x_1$ and $x_2$, are convolved with a Gaussian kernel to produce smoothed, continuous firing rates. As the kernel bandwidth increases, the correlation between the smoothed spike trains also increases. However, this observed correlation reflects both the true underlying correlation and an artifactual component introduced by the smoothing process. To disentangle this effect, the original spike trains are randomly permuted to generate surrogate spike trains, $\tilde{x}_1$ and $\tilde{x}_2$ (surrogate), which are uncorrelated by design. These surrogates are then smoothed using the same Gaussian kernel. Notably, the correlation between the smoothed surrogate spike trains also increases with kernel bandwidth, despite the absence of intrinsic correlation, indicating that this increase is solely due to smoothing. The "Informative Correlation," shown as a black bar, is defined as the difference between the correlation of the original smoothed spike trains and that of the smoothed surrogates, thereby isolating the portion of the correlation attributable to genuine spike train similarity rather than smoothing artifacts.}
\label{Fig:surrogated baseline problem}
\end{figure}

However, several factors in this approach require further consideration due to their significant impact on the interpretation of the results. First, the smoothing kernel bandwidth, which determines the degree of smoothness in firing rate estimation, is often empirically selected. Kernel bandwidth plays a crucial role in calculating the correlations between spike trains. As shown in Figure \ref{Fig:surrogated baseline problem}, if the bandwidth is too narrow, it fails to capture true correlations due to the sparsity of the spike trains. Conversely, an overly broad bandwidth can artificially inflate correlations and even introduce spurious relationships between uncorrelated spike trains. Therefore, optimizing the kernel bandwidth is essential for accurately quantifying the similarity between spatio-temporal spike patterns.

Second, baseline correlations arising from the stochastic nature of spike firing must be accounted for. Even randomly generated spike trains can exhibit correlations within a short time window, and these correlations tend to increase with kernel width (gray bars in Figure \ref{Fig:surrogated baseline problem}). This spurious correlation introduces a nonzero baseline that biases similarity estimates, particularly for large kernels. To ensure accurate similarity measurements, such baseline correlations should be subtracted.

\begin{figure}[!htbp]
\begin{center}
\includegraphics[width=.9\textwidth]{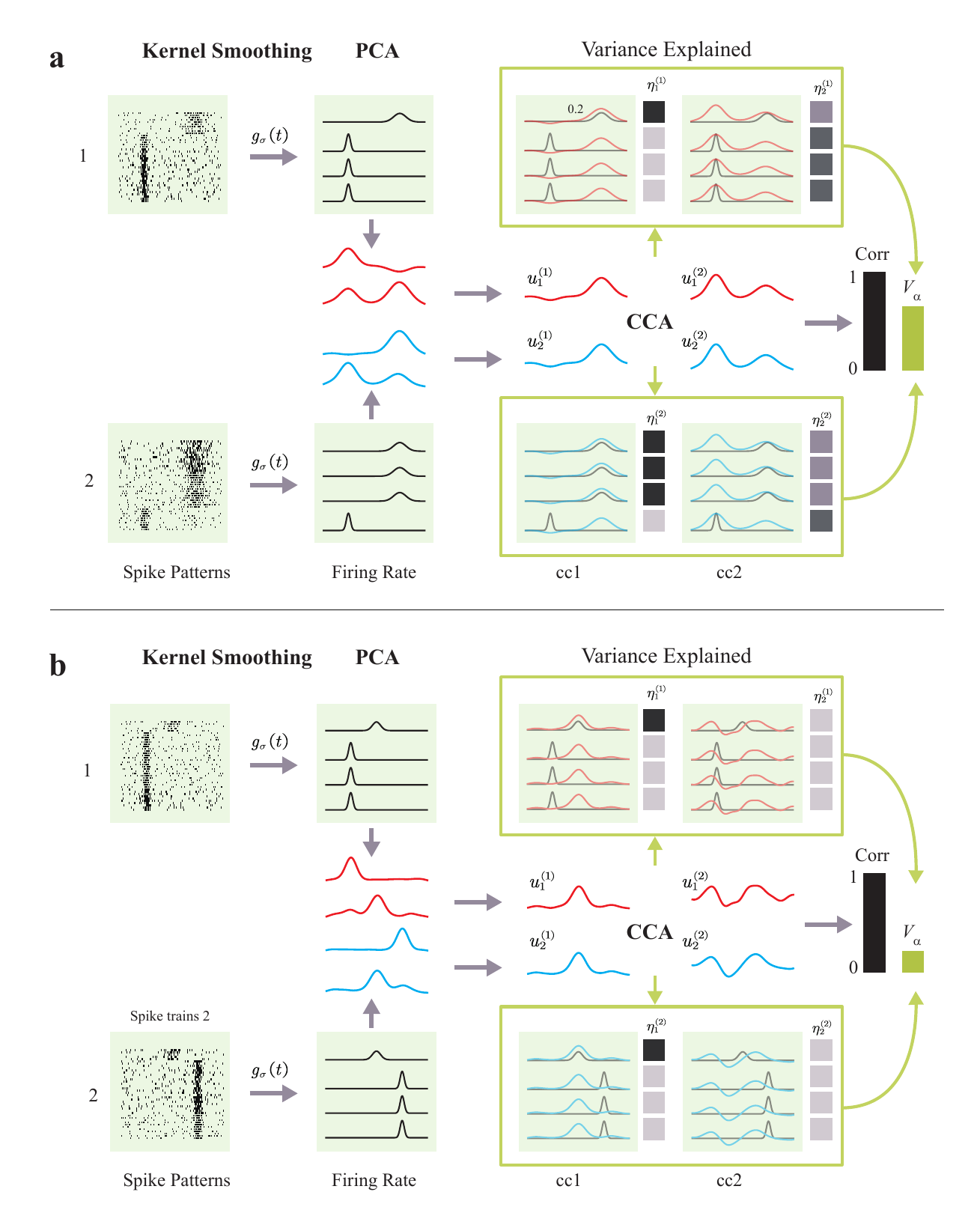}
\end{center}
\caption{Using PCA and CCA to identify and align latent dynamics. \textbf{a.} Two spike matrices share the same underlying latent dynamics but consist of different subsets of neurons. PCA is first applied to extract latent dynamics from each pattern, followed by CCA to align them. While CCA successfully identifies the shared structure and treats the aligned components as similar, it does not consider the proportion of variance explained by each canonical component (e.g., cc1 or cc2). \textbf{a.} In contrast, two spike matrices exhibit largely distinct latent dynamics, with only a small subset of neurons sharing similar activity patterns. CCA disproportionately emphasizes these shared dynamics, resulting in aligned components that capture only a small fraction of the total variance, as most of the structure in the original spike patterns differs.}
\label{Fig:cca problem}
\end{figure}

Third, using CCA to align latent dynamics and quantify pattern similarity overlooks the extent to which latent dynamics explain variance in the original spike patterns. While PCA prioritizes components that capture the most variance, CCA applies an additional linear transformation that maximizes the correlation between latent dynamics, regardless of their explanatory power in the original spike patterns. As a result, this method can yield high similarity scores even for very different spike patterns. For example, as shown in Figure \ref{Fig:cca problem}\textbf{a}, two spike patterns generated from the same set of latent dynamics will always be considered identical despite significant differences in the variance each dynamic explains in each spike pattern. More critically, CCA may ignore important latent dynamics in the spike patterns by prioritizing those that maximize correlations across patterns (Figure \ref{Fig:cca problem}\textbf{b}). Although PCA can partially mitigate this issue by prescreening less significant latent dynamics, it does not account for their relative importance in explaining variance within the original spike patterns.

To address the limitations of existing approaches, we propose Relevance-Balanced Continuum Correlation Analysis by smoothing and surrogating (ReBaCCA-ss), a novel method for robustly quantifying the similarity between spatio-temporal patterns of spikes in consideration of the variance explained, while simultaneously identifying the optimal kernel bandwidth for this similarity measure. Our approach introduces several key innovations:

\begin{itemize}
    \item \textbf{Unified Framework for Alignment and Variance Explanation.} 
    
    Inspired by the continuum regression framework \citep{lee2007continuum, xie2020optimizing}, ReBaCCA-ss unifies cross-pattern alignment and within-pattern variance explanation within a single framework controlled by a tunable parameter $\alpha\in[0,1]$. This parameter balances two objectives: maximizing canonical correlation (alignment strength) and prioritizing dimensions that explain significant variance in the original spike patterns.
    
    For a given $\alpha$, the method constructs a projection space that aligns the latent dynamics of two spike patterns. Each aligned dimension is weighted according to its pattern-specific variance explained. The weighted sum of canonical correlations is termed Relevance-Balanced Continuum Correlation Analysis (ReBaCCA).
    
    \item \textbf{Surrogate-Based Noise Correction.} 
    
    To correct for noise, we generate surrogate spiking data by permuting spike timing while preserving firing rates, thereby disrupting temporal correlations. The ReBaCCA calculated from these surrogate patterns quantifies baseline noise correlation, which is then subtracted from the ReBaCCA of the original spike patterns to isolate the true similarity. This corrected metric, termed ReBaCCA-ss, ensures robustness against the stochastic nature of neural firing.
    
    \item \textbf{Optimal Kernel Bandwidth Selection}
    
    ReBaCCA-ss is computed across a range of kernel bandwidths. For each bandwidth, the method evaluates the separation between the real data correlation and the surrogate data correlation. The optimal bandwidth is selected as the one that maximizes this separation, providing a data-driven approach for optimizing kernel bandwidth in the context of spike pattern similarity analysis.
    
\end{itemize}

We validate ReBaCCA-ss using both simulated data and experimental data recorded from rats performing a Delayed Nonmatch-to-Sample (DNMS) task \citep{song2009nonlinear,song2014extraction}. Simulation results demonstrate the limitations of existing methods and show that our approach effectively balances alignment and variance explanation while, accurately identifying the optimal kernel bandwidth for reconstructing ground-truth latent dynamics. With experimental data, ReBaCCA-ss successfully identifies the optimal kernel bandwidth and quantifies similarity between spike patterns. In combination with Multidimensional Scaling (MDS) \citep{carroll1998multidimensional}, pair-wise ReBaCCA-ss between all spike patterns reveal how events are represented within the same and across different trials, sessions, and animals. This framework provides a powerful tool for investigating how information is represented in neuronal spiking activities.

\newpage
\section{Methods}

\subsection{Continuum Canonical Correlation: Bridging PCA and CCA}

First, we determine a low-dimensional projection space that considers both the linear correlation between the two patterns (as in CCA) and their individual variances (as in PCA). Continuum Regression (CR) offers a way to balance correlation and variance in regression problems controlled by a tuning parameter $\alpha \in [0,1]$, which determines the trade-off between these two objectives \citep{stone1990continuum,sundberg1993continuum,bjorkstrom1999generalized}. For small values of $\alpha$, CR emphasizes correlation after regression, while for larger values, it focuses more on variance. However, CR is inherently a multiple-to-one problem, making it unsuitable for directly comparing two spike train matrices. To address this, a more general approach for multivariate data termed Continuum Canonical Correlation (CCC) can be used \citep{lee2007continuum}. Denote two spike patterns as matrices $X_1\in \mathbb{R}^{\mathcal{T}\times N_1}$ and $X_2\in \mathbb{R}^{\mathcal{T}\times N_2}$, where $\mathcal{T}$ represents the trial time, and $N_1$ and $N_2$ denote the number of neurons in each spike train, respectively. Each element in $X_1$ and $X_2$ is binary, with 0 or 1 indicating the absence or presence of a spike at a specific time for a given neuron. CCC identifies the projection vectors $w_1\in \mathbb{R}^{N_1\times 1}$ and $w_2\in \mathbb{R}^{N_2\times 1}$ for each spike matrix that maximizes the following objective function.
\begin{align}
    \underset{w_1,w_2}{\operatorname{max}}\quad & \Bigl(\frac{w_1^TX_1^TX_1w_1}{\text{trace}(X_1^TX_1)}\Bigl)^\alpha \Bigl(\frac{(w_1^TX_1^TX_2w_2)^2}{w_1^TX_1^TX_1w_1\cdot w_2^TX_2^TX_2w_2}\Bigl)^{1-\alpha} \Bigl(\frac{w_2^TX_2^TX_2w_2}{\text{trace}(X_2^TX_2)}\Bigl)^\alpha 
    \label{eq:obj_fun}
\end{align}
The left and right components represent the variance explained after projecting the two spike matrices, respectively, each raised to the power of $\alpha$. The middle component is the squared correlation between the two projected vectors, raised to the power of $1-\alpha$. The objective is to achieve a balance between variance explained and correlation, controlled by $\alpha$. An exact balance is reached when $\alpha=0.5$.

\begin{figure}[!htbp]
\centering
\includegraphics[width=\textwidth]{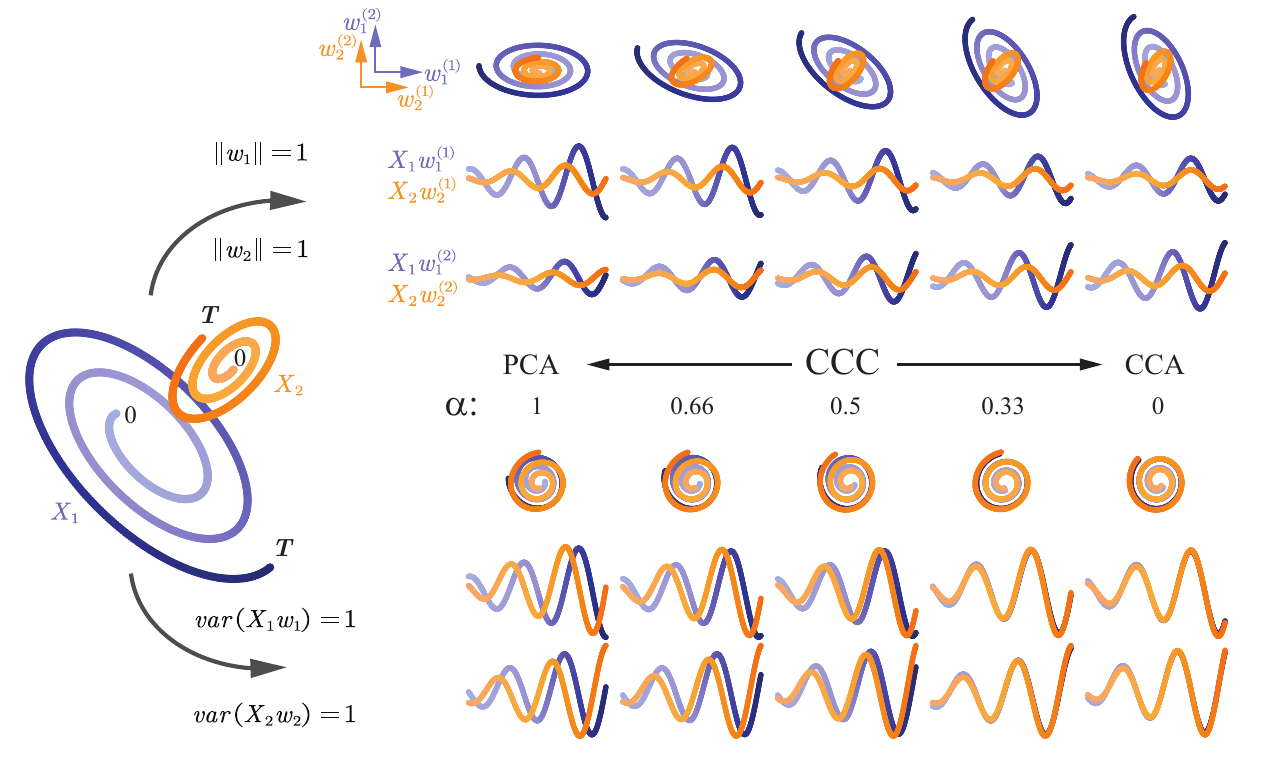}
\caption{CCC of two temporal patterns under different $\alpha$. The patterns to be aligned are represented by orange and blue spirals, respectively. The color gradient from light to dark indicates the temporal progression of the data points. The first three rows present the CCC outcomes with the projection vectors' norm set to 1, same as PCA constraints. The first row displays the 2D projections, while the second and third rows show the temporal evolution of the projection dimension. The last three rows demonstrate the impact of $\alpha$ under CCA-like constraints, where the total variance of the projections is set to 1.}
\label{Fig:CCC visualization}
\end{figure}

Figure \ref{Fig:CCC visualization} illustrates the effect of the tuning parameter $\alpha$ on the CCC transformation of two patterns $X_1$ and $X_2$. For $\alpha=1$, CCC maximizes the variance of each dataset independently, yielding PCA-like results. This approach prioritizes variance maximization, effectively aligning each pattern along its direction of largest variance without considering inter-dataset correlations. Consequently, the blue and orange curves exhibit minimal correlation, with misaligned peaks. As $\alpha$ decreases, CCC increasingly emphasizes the correlation between the projected patterns. This is visually represented by the clockwise rotation of the blue spiral and the counterclockwise rotation of the orange spiral. When $\alpha=0$, the objective shifts entirely to maximizing the correlation, disregarding variance. Consequently, the projections align perfectly, achieving a correlation of 1. While the PCA-like constraints facilitate visualization of rotational effects, the CCA-like constraints provide a more direct illustration of correlation changes (last two rows). As $\alpha$ decreases from 1 to 0, the emphasis on correlation increases, leading to a progressive rise in the correlation coefficient from 0 to 1.

\subsection{Relevance Balanced Continuum Correlation Analysis (ReBaCCA)}

\begin{figure}[!htbp]
\centering
\includegraphics[width=\textwidth]{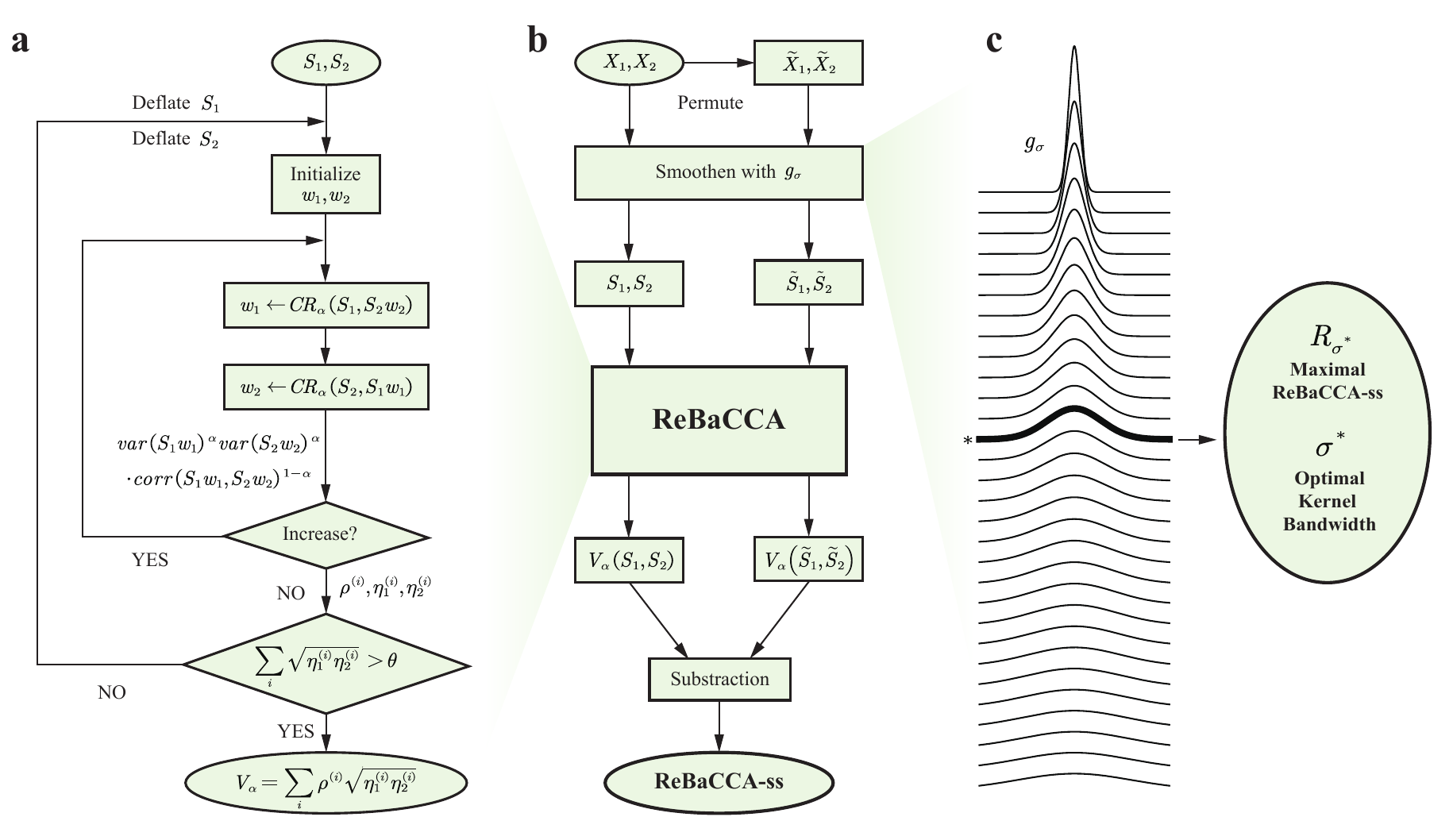}
\caption{The ReBaCCA framework. \textbf{a.} Relevance Balanced Continuum Correlation Analysis (ReBaCCA). \textbf{b.} ReBaCCA with smoothing and surrogating (ReBaCCA-ss). \textbf{c.} ReBaCCA-ss with optimal kernel bandwidth.}
\label{Fig:algorithm flowchart}
\end{figure}

While CCC identifies the projection directions for each pattern, previous work \citep{lee2007continuum} did not evaluate the overall relationship between the patterns. In this study, we extend CCC by incorporating the variance explained after projection as a weighting factor. We then integrate these weighted correlations across all projection dimensions to form a single similarity metric, which we call Relevance Balanced Continuum Correlation Analysis (ReBaCCA). A detailed flowchart of this approach is provided in Figure \ref{Fig:algorithm flowchart}\textbf{a}. The method is implemented as follows.

For two spike train patterns $X_1$ and $X_2$, given a specific Gaussian kernel $g_{\sigma}(t) = \frac{1}{\sigma\sqrt{2\pi}} e^{-\frac{t^2}{2\sigma^2}}$, where $\sigma$ is the kernel bandwidth, both spike trains are smoothed as follows:
\begin{align}
    & S_1=X_1*g_\sigma(t) \\
    & S_2=X_2*g_\sigma(t)
\end{align}
Here the $*$ symbol denotes convolution, meaning that each column of the matrices $X_1$ and $X_2$ is convolved with the same Gaussian kernel $g_\sigma(t)$. Then $S_1$ and $S_2$ are provided as inputs to the CCC algorithm. To optimize $w_1$ and $w_2$ with the objective function in (\ref{eq:obj_fun}), we employ an iterative procedure: first, we fix one vector (either $w_1$ or $w_2$) and update the other vector by maximizing the objective function. Then we fix the updated vector and update the other vector. This process repeats until convergence. The initial values of $w_1$ and $w_2$ are set empirically as follows: if $\alpha\leq 0.5$, $w_1$ and $w_2$ are the first projection vectors of $\text{CCA}(S_1,S_2)$. If $\alpha>0.5$,  $w_1$ and $w_2$ are the first projection vectors of $\text{PCA}(S_1)$ and $\text{PCA}(S_2)$, respectively. 

In each step, when one vector is fixed, the problem reduces to a continuum regression between $S_1$ and $S_2w_2$. The solution at each step is \citep{xie2020optimizing}:
\begin{align}
    & w_1^*\propto (S_1^TS_1)^{(\frac{\alpha}{1-\alpha}-1)}S_1^TS_2w_2 \label{eq:w1}\\
    & ||w_1^*|| = 1
\end{align}

Given the current optimal $w_1$, the solution for $w_2$ is 
\begin{align}
    & w_2^*\propto (S_2^TS_2)^{(\frac{\alpha}{1-\alpha}-1)}S_2^TS_1w_1 \\
    & ||w_2^*|| =1 \label{eq:w2}
\end{align}

We repeat (\ref{eq:w1}-\ref{eq:w2}) to find the first pair of projection vectors that maximize the cost function. We denote them as $w_1^{(i)^*}$ and $w_2^{(i)^*}$ ($i=1$). The current smoothed spike trains are denoted as $S_1^{(i)}$ and $S_2^{(i)}$, where $S_1^{(1)}=S_1$ and $S_2^{(1)}=S_2$. The projections of the aligned vectors in dimension $i$ are:
\begin{align}
    &u^{(i)}_1=S_1^{(i)}w^{(i)^*}_1 \\
    &u^{(i)}_2=S_2^{(i)}w^{(i)^*}_2
\end{align}
The variances explained for $S_1$ and $S_2$ are denoted as
\begin{align}
&\eta^{(i)}_1=\dfrac{(u^{(i)}_1)^Tu^{(i)}_1}{\text{tr}(S_1^TS_1)}\\ 
&\eta^{(i)}_2=\dfrac{(u^{(i)}_2)^Tu^{(i)}_2}{\text{tr}(S_2^TS_2)} 
\end{align}
Here, $\text{tr}(\cdot)$ represents the trace of the matrix. The joint variance explained in dimension $i$ from both $S_1$ and $S_2$ is denoted as the geometric mean of variance explained from two datasets 
\begin{equation}
   \rho^{(i)}=\sqrt{\eta^{(i)}_1 \eta^{(i)}_2}
\end{equation}

Next, we perform matrix deflation as:
\begin{align}
&  S_1^{(i+1)}=S_1^{(i)}-u^{(i)}_1\Bigl(\frac{S_1^{(i)^T}u^{(i)}_1}{u_1^{(i)^T}u_1^{(i)}}\Bigr)^T\\
&  S_2^{(i+1)}=S_2^{(i)}-u^{(i)}_2\Bigl(\frac{S_2^{(i)^T}u^{(i)}_2}{u_2^{(i)^T}u_2^{(i)}}\Bigr)^T
\end{align}

For the pair of deflated matrices $S_1^{(i+1)}$ and $S_2^{(i+1)}$, we repeat (\ref{eq:w1}-\ref{eq:w2}) to find the projection vectors $w_1^{(i+1)^*}$ and $w_2^{(i+1)^*}$. This procedure continues until the total variance explained exceeds a threshold $\theta$ ($N_{\theta}=\min{\{N: \sum_{i=1}^{N}\rho^{(i)}>\theta\}}$), as shown in the outer loop in Figure \ref{Fig:algorithm flowchart}\textbf{a}. The aligned latent dynamics are denoted as
\begin{align}
& L_1=[u^{(1)}_1,u^{(2)}_1,...,u^{(N_{\theta})}_1] \\
& L_2=[u^{(1)}_2,u^{(2)}_2,...,u^{(N_{\theta})}_2] 
\end{align}

The final metric ReBaCCA that quantifies the similarity between $S_1$ and $S_2$ is:
\begin{equation}
    V_{\alpha}(S_1,S_2)=\sum_{i=1}^{N_{\theta}}\rho^{(i)} \cdot \text{corr}(u^{(i)}_1,u^{(i)}_2) \label{eq:rebacca}
\end{equation}
Given that the sum of the variance explained by each spike pattern is less than or equal to 1 ($\sum_i^{N_{\theta}}\eta_1^{(i)}\leq1$ and $\sum_i^{N_{\theta}}\eta_2^{(i)}\leq1$), based on Cauchy-Schwarz inequality, the sum of $\rho^{(i)}$ is between 0 and 1 ($0\leq\sum_i^{N_{\theta}}\rho^{(i)}\leq1$). Furthermore, after alignment, the correlation between the projected variables $u_1^{(i)}$ and $u_2^{(i)}$ also ranges between 0 and 1. Hence, the upper bound of the ReBaCCA value $V_{\alpha}(S_1,S_2)$ is given by:
\begin{equation}
    V_{\alpha}(S_1,S_2)=\sum_{i=1}^{N_{\theta}}\rho^{(i)} \cdot \text{corr}(u^{(i)}_1,u^{(i)}_2)\leq \sum_{i=1}^{N}\rho^{(i)} \leq 1 \label{eq:rebacca range}
\end{equation}
Consequently, the ReBaCCA value always ranges between 0 and 1, providing a bounded, quantitative measure of similarity between two spike patterns.

The connection between the ReBaCCA computation and the underlying objective function (\ref{eq:obj_fun}) is as follows. When setting $\alpha=0.5$, Equation (\ref{eq:rebacca}) is equivalent to summing the objective values across all $N$ aligned dimensions. For subsequent analysis in this paper, we set $\alpha=0.5$, ensuring a natural balance between explained variance and aligned correlation, and maintaining a consistent objective function range within [0, 1].

The choice of $\alpha$ significantly impacts the scaling of the objective function because each explained variance term is raised to the power of $2\alpha$.  If $\alpha\geq0.5$, then $\sum_i^{N_{\theta}}\eta_1^{(i)^{2\alpha}}\leq1,\sum_i^{N_{\theta}}\eta_2^{(i)^{2\alpha}}\leq1$, the sum of objective functions is between 0 and 1. If $\alpha<0.5$, then $\sum_i^{N_{\theta}}\eta_1^{(i)^{2\alpha}}\geq1$ and $\sum_i^{N_{\theta}}\eta_2^{(i)^{2\alpha}}\geq1$. In this case, the sum of ${N_{\theta}}$ objective functions is not guaranteed to be within 0 and 1. To avoid this scale inconsistency with different $\alpha$, a two-step strategy is adopted to calculate ReBaCCA. First, $\alpha$ is used to determine the optimal projection space. Once this space is determined, the overall similarity can be evaluated using Equation (\ref{eq:rebacca}) without involving $\alpha$.

The proposed ReBaCCA value offers a more robust quantification of spike pattern similarity than commonly used CCA-based approaches. Traditional CCA does not yield a single, universally accepted similarity metric. The first approach relies on the maximum canonical correlation—the correlation between the first pair of canonical variables—as the similarity metric \citep{marek2022reproducible}. However, this approach captures only the most dominant aligned component and ignores contributions from the remaining dimensions. A second approach averages the correlations across a fixed number of top canonical variables (e.g., the top six) \citep{gallego2020long}. While this method incorporates more dimensions, the choice of how many to include can influence the similarity evaluation considerably: using fewer dimensions may inflate the similarity score, while including more may dilute it.

In contrast, the ReBaCCA framework addresses these limitations by weighting each aligned dimension according to the variance it explains in the original data. This ensures that all aligned components contribute to the final similarity score in proportion to their relevance. By aggregating this weighted information, ReBaCCA provides a more comprehensive measure of similarity between population spike patterns. The pseudocode for computing the ReBaCCA value is provided in Algorithm \ref{alg:pseudo rebacca}.

\begin{algorithm}[!htbp]
\caption{Pseudo Code for ReBaCCA}
\label{alg:pseudo rebacca}
\KwIn{$S_1\in \mathbb{R}^{\mathcal{T}\times N_1}, S_2\in \mathbb{R}^{\mathcal{T}\times N_2}$}
\KwOut{Relevance Balanced Continuum Canonical Analysis $V_{\alpha}$}
\SetKwBlock{Begin}{Function ReBaCCA($S_1, S_2$):}{end}

\textbf{Parameters:} \\
\quad $\alpha\in[0,1]:$ Variance-correlation trade-off \\
\quad $M:$ Maximal number of iterations for continuum canonical correlation \\
\quad $\theta:$ Variance explained threshold

\Begin{
    \textbf{Initialize} $S_1^{(1)} \gets S_1, \, S_2^{(1)} \gets S_2, \, f^{(0)} = 0$ \;
    \For{$i \gets 1$ \KwTo $\min(N_1,N_2)$}{
        Initialize $w_1^{(i)}$ and $w_2^{(i)}$ \;
        \For{$j \gets 1$ \KwTo $M$}{
            $w_1^{(i)} \gets \mathrm{CR}_\alpha (S_1^{(i)}, \, S_2^{(i)} w_2^{(i)})$ \;
            $w_2^{(i)} \gets \mathrm{CR}_\alpha (S_2^{(i)}, \, S_1^{(i)} w_1^{(i)})$ \;
            $f^{(i)} \gets \mathrm{Loss}(S_1^{(i)}w_1^{(i)},S_2^{(i)}w_2^{(i)}) $ Equation (\ref{eq:obj_fun}) \; 
            \If{$f^{(i)} \le f^{(i-1)}$}{
               \textbf{break}\;
            }
        }
        $V^{(i)} \gets \rho^{(i)} \cdot \mathrm{corr}(S_1^{(i)}w_1^{(i)},S_2^{(i)}w_2^{(i)})$ \;
        \If{$\sum \rho^{(i)} > \theta$}{
            \textbf{break}\;
        }
        \Else{
            $S_1^{(i+1)} \gets \text{Deflate} \ S_1^{(i)}$ \;
            $S_2^{(i+1)} \gets \text{Deflate} \ S_2^{(i)}$ \;
        }
    }
    \Return{$V_{\alpha} = \sum V^{(i)}$}\;
}
\end{algorithm}

\subsection{Relevance Balanced Canonical Correlation Analysis with Smoothing and Surrogating (ReBaCCA-ss)}

\begin{figure}[!htbp]
\centering
\includegraphics[width=0.5\textwidth]{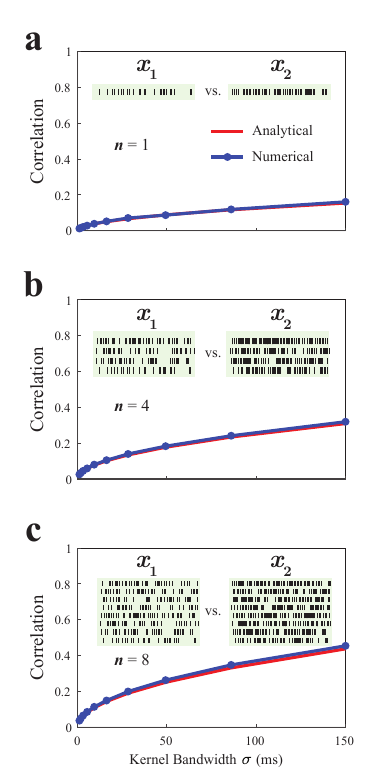}
\caption{Relationship between CCA correlation, kernel bandwidth, and number of neurons for independently generated spike train matrices. The red curve represents the analytical correlation value, while the blue curve depicts the average correlation derived from 500 simulation trials. \textbf{a.} 1 neuron vs 1 neuron. \textbf{b.} 4 neuron vs 4 neuron. \textbf{c.} 8 neuron vs 8 neuron.}
\label{Fig:cca theoretical corr}
\end{figure}

In prior subsections, we introduced ReBaCCA as a method for analyzing similarity between two patterns, yet it does not address the correlation artifacts introduced solely by kernel smoothing in spike data. This subsection first delves into the phenomenon of kernel smoothing-induced correlation in CCA, even when applied to independent Poisson spike train matrices. Consider two independent spike train matrices, each consisting of $N$ neurons and recorded over a trial duration $T$. Assuming that the firing rate of each neuron is low relative to $T$, and that the kernel bandwidth $\sigma$ is small compared to $T$, the average CCA correlation after Gaussian kernel smoothing with $g_\sigma(t)$ is expressed as (a comprehensive proof is provided in the Appendix):
\begin{equation}
    (\frac{8}{\pi})^{1/4}\cdot\sqrt{\frac{N\sigma}{T}}
\end{equation}
This formula captures the intuitive correlation coefficient outcomes for CCA between two independent, smoothed spike train matrices. As depicted in Figure \ref{Fig:cca theoretical corr}, the correlation increases with the total number of neurons $N$. This effect occurs because a larger $N$ provides CCA with more components for linear combinations, thereby amplifying the potential to produce correlated variables. Similarly, the correlation rises with the kernel bandwidth $\sigma$, as wider kernels increase the similarity between spike trains. Additionally, this limitation of CCA is influenced by the finite recording duration $T$ of each trial. When $T$ is sufficiently large, the kernel smoothing-induced correlation becomes negligible; however, this condition is rarely satisfied in practical experimental contexts. Consequently, this effect must be considered when evaluating spike pattern similarity. To address this, our proposed ReBaCCA-ss method employs surrogate spike matrices as a baseline to quantify the similarity induced by kernel smoothing.

For each spike train matrix $X_1$ and $X_2$, we define the permutation matrices $ P_1,P_2 \in \mathbb{R}^{\mathcal{T} \times \mathcal{T}} $, which are binary matrices where each row and each column contains exactly one entry equal to 1, and all other entries are 0. The surrogate spike train matrices are then generated as follows:
\begin{align}
    & \tilde{X}_1=P_1X_1 \\
    & \tilde{X}_2=P_2X_2
\end{align}

As shown in Figure \ref{Fig:algorithm flowchart}\textbf{b}, we apply the same Gaussian kernel function to smooth the surrogate spike trains:
\begin{align}
    & \tilde{S}_1=\tilde{X}_1*g_\sigma(t) \\
    & \tilde{S}_2=\tilde{X}_2*g_\sigma(t)
\end{align}
Since the surrogate spike trains are generated randomly, there should be no inherent correlation between them. Thus, any similarity measure between $\tilde{S}_1$ and $\tilde{S}_2$ is expected to originate solely from the effect of the kernel smoothing. We introduce a new similarity measure called Relevance Balanced Canonical Correlation Analysis with Smoothing and Surrogating (ReBaCCA-ss) to account for this effect:
\begin{equation}
    R_{\sigma|\alpha}(X_1,X_2)=V_{\alpha}(S_1,S_2)-V_{\alpha}(\tilde{S}_1, \tilde{S}_2)
\end{equation}

\subsection{ReBaCCA-ss with optimal kernel bandwidth}

Given two spike matrices, $X_1$ and $X_2$, the choice of kernel bandwidth $\sigma$ significantly affects the similarity between their smoothed representations. When the kernel bandwidth is small $\sigma\to 0$, the similarities between $S_1$ and $S_2$, as well as between $\tilde{S}_1$ and $\tilde{S}_2$, are both low. This occurs due to the inherent sparseness in the spike timings. In this case, the ReBaCCA-ss measure is small.

As the kernel bandwidth increases, the similarity between $S_1$ and $S_2$, as well as between $\Tilde{S}_1$ and $\Tilde{S}_2$, also increases, but at different rates. The ReBaCCA of the original spike trains rises more rapidly when they are correlated. In contrast, the correlation between the permuted spike trains grows at a slower rate, which reflects the impact of the kernel bandwidth on the correlation. To accurately quantify the relationship between the two spike matrices, we determine the optimal kernel bandwidth as the one that maximizes the difference between the similarity of the original smoothed spike trains and that of the surrogate smoothed spike trains as:
\begin{equation}
    \underset{\sigma}{\operatorname{maximize}}\quad R_{\sigma|\alpha}(X_1,X_2)
\end{equation}
The optimal value $R_{\sigma^*|\alpha}$ describes the similarity between the two spike train matrices with the optimal kernel width $\sigma^*$, which is applied for the subsequent analysis. The pseudo-code of calculating $R_{\sigma^*|\alpha}$ is shown in Algorithm \ref{alg:pseudo rebacca-ssok}.

\begin{algorithm}[!htbp]
\caption{Pseudo Code for ReBaCCA-ssok}
\label{alg:pseudo rebacca-ssok}

\KwIn{$X_1 \in \mathbb{R}^{\mathcal{T}\times N_1}, \quad X_2 \in \mathbb{R}^{\mathcal{T}\times N_2}$}
\KwOut{Optimal kernel bandwidth $\sigma^*$, Informative correlation $R_{\sigma^*|\alpha}$}

\textbf{Parameters:} \\
\quad $\alpha$: continuum parameter \\
\quad $\boldsymbol{\sigma} = (\sigma_1,\sigma_2,\dots,\sigma_n)$: kernel width pool \\
\quad $P_1, P_2$: permutation matrices

\SetKwBlock{Begin}{\textbf{Function} $\text{ReBaCCA-ssok}(X_1,X_2)$:}{end}
\Begin{
    Generate permuted matrices $\widetilde{X}_1 \leftarrow P_1 X_1,\quad \widetilde{X}_2 \leftarrow P_2 X_2$\; 
    \For{$\sigma_i \in \boldsymbol{\sigma}$}{
        $S_1, S_2 \leftarrow$ Smooth the matrices $X_1$ and $X_2$ using Gaussian kernel $g_{\sigma_i}$\;
        $\widetilde{S}_1, \widetilde{S}_2 \leftarrow$ Smooth the matrices $\widetilde{X}_1$ and $\widetilde{X}_2$ using Gaussian kernel $g_{\sigma_i}$\;
        Calculate $V_{\alpha}(S_1, S_2)$\;
        Calculate $V_{\alpha}(\widetilde{S}_1, \widetilde{S}_2)$\;
        Compute $R_{\sigma_i|\alpha} = V_{\alpha}(S_1, S_2) \;-\; V_{\alpha}(\widetilde{S}_1, \widetilde{S}_2)$\;
    }
    $\sigma^* \leftarrow \arg\max_{\sigma_i} R_{\sigma_i|\alpha}$\;
    \Return $(\sigma^*,\,R_{\sigma^*|\alpha})$\;
}

\end{algorithm}

\newpage
\section{Simulation Results}

\begin{figure}[!htbp]
\begin{center}
\includegraphics[width=\textwidth]{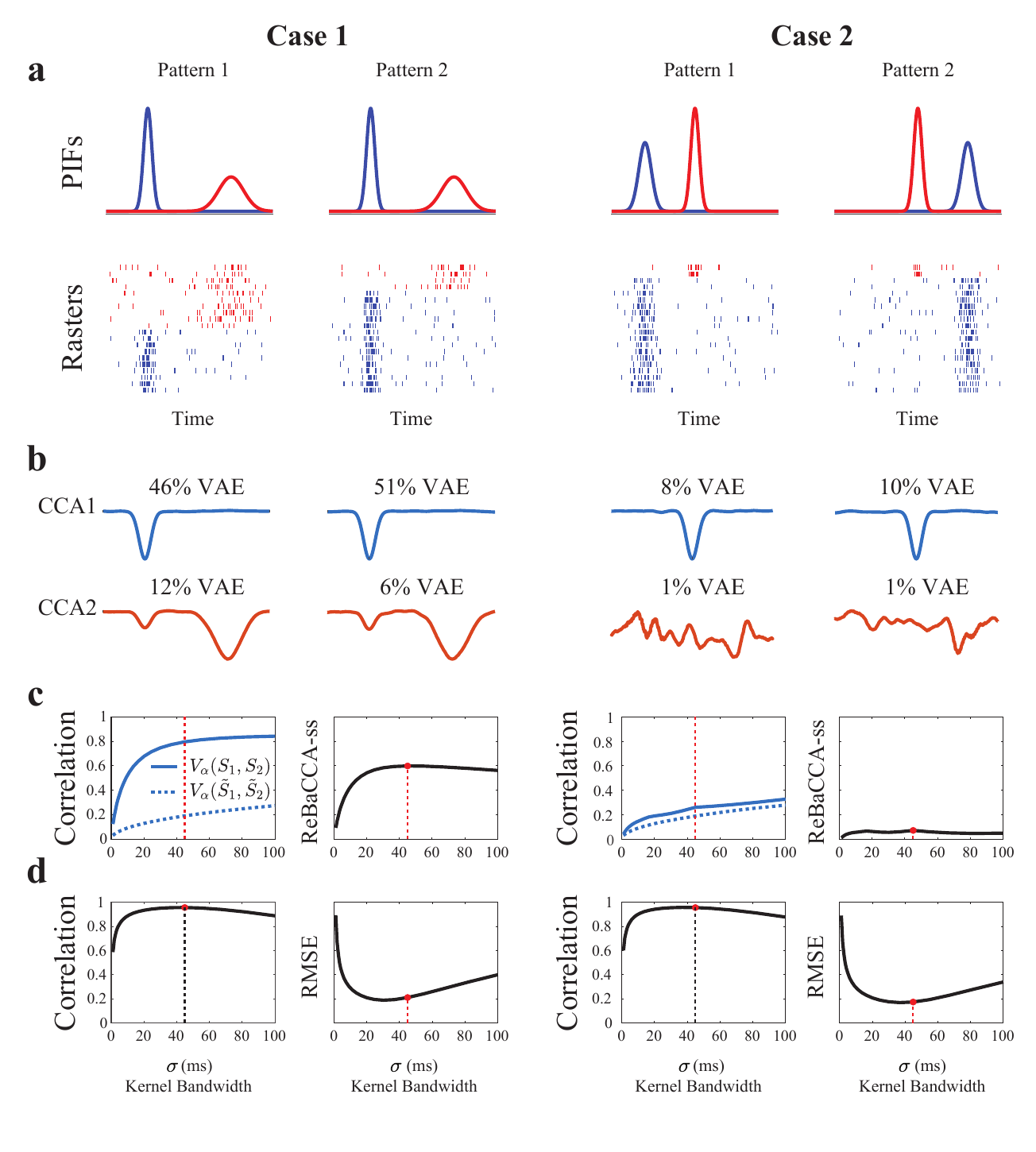}
\end{center}
\caption{Calculating ReBaCCA and optimal kernel bandwidth between simulated spike patterns. \textbf{a.} Firing intensity functions and spike rasters of two groups of neurons. Each group has two temporal firing patterns represented by the blue and red curves. \textbf{b.} First (blue) and second (red) aligned dimensions obtained after kernel smoothing, PCA, and CCA. \textbf{c.} ReBaCCA values for the original smoothed spike patterns (solid blue curve) and for surrogated spike patterns (dashed blue curve) as a function of kernel bandwidth. The black curve represents the ReBaCCA-ss value, with the red dot indicating the optimal kernel bandwidth that maximizes ReBaCCA-ss. \textbf{d.} Correlation and root-mean-squared error (RMSE) between the reconstructed latent dynamics and the ground truth, as a function of kernel width. The red dot marks the performance at the optimal kernel width.}
\label{Fig:simulated setting}
\end{figure}

To evaluate our proposed methods, we generated simulated data for two groups of neurons under different scenarios (Figure \ref{Fig:simulated setting}\textbf{a}), where each group of neurons exhibits two distinct temporal firing patterns. In Case 1, both groups share identical temporal firing patterns, but the proportion of neurons displaying each pattern varies between the groups. In Case 2, the majority of neurons in each group exhibit distinct firing patterns (blue curves), while a small subset shares similar firing patterns (red curves). In both cases, CCA aligns the two datasets with near-perfect correlation due to the presence of shared patterns (Figure \ref{Fig:simulated setting}\textbf{b}). However, this high correlation is misleading, as the two spike patterns differ significantly. This discrepancy arises because CCA maximizes correlation through linear combinations without accounting for the variance explained by each aligned dimension in the original data (e.g., the aligned latent dynamics only have 8\% and 10\% VAE in Case 2, Figure \ref{Fig:simulated setting}\textbf{b}).

In contrast, our proposed ReBaCCA method integrates both the aligned correlation and the variance explained (VAE) in each dataset ($\alpha=0.5$). The resulting ReBaCCA values across different kernel widths are presented in Figure \ref{Fig:simulated setting}\textbf{c}. In Case 1, where the original spike patterns are similar but differ in the proportion of neurons, the ReBaCCA values for the original smoothed matrices (solid blue curve) increase more rapidly with kernel width compared to those for the surrogate smoothed matrices (dashed blue curve). The ReBaCCA-ss value peaks at approximately 0.6 at an optimal kernel width of 45 ms. In Case 2, where the majority of spike patterns are dissimilar, the ReBaCCA values for the original smoothed matrices are only slightly higher than those for the surrogate matrices. Here, the ReBaCCA-ss value reaches a maximum of approximately 0.05, also at a kernel width of 45 ms. These simulation results underscore the strengths of our method, which jointly considers VAE and aligned correlation. For datasets with similar latent dynamics but differing proportions, our approach assigns a relatively high similarity score without suggesting complete identity. Conversely, when only a small portion of the data exhibits similarity—--a feature that CCA might overemphasize—--our method yields a low similarity score, comparable to that of surrogate matrices.

Furthermore, after determining the optimal kernel width using $R_{\sigma^*|\alpha}$, we employed PCA to reconstruct the latent dynamics from the smoothed spike matrices. As illustrated in Figure \ref{Fig:simulated setting}\textbf{d}, the correlation between the reconstructed and ground truth latent dynamics is very high. While the root-mean-squared error (RMSE) does not reach its minimum, it remains within a relatively small range. These findings indicate that the optimal kernel width identified by our method effectively captures the underlying latent dynamics of the spike patterns.

\begin{figure}[!htbp]
\begin{center}
\includegraphics[width=\textwidth]{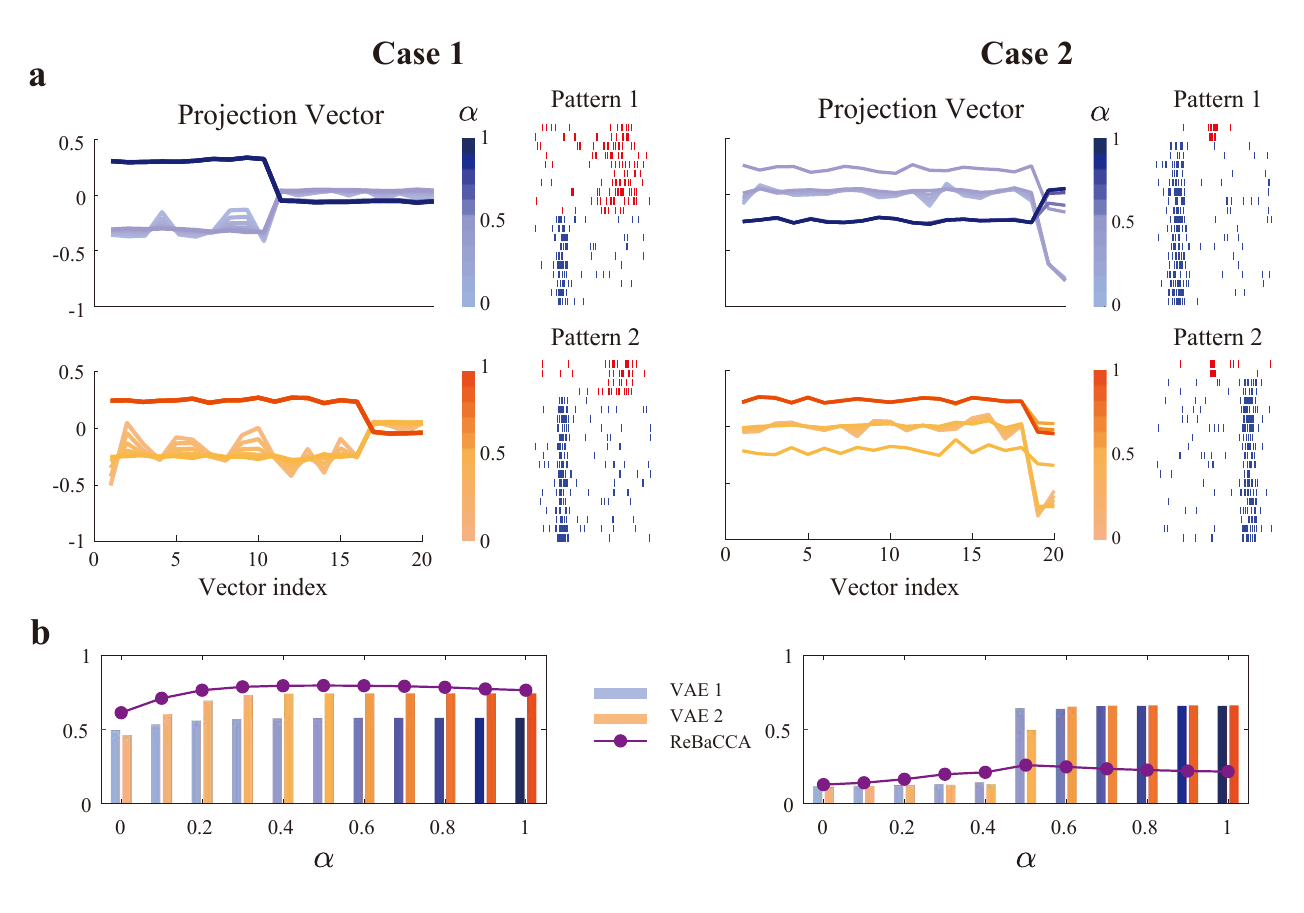}
\end{center}
\caption{\textbf{a.} Projection vectors of the first aligned dimension of the two spike patterns. The x-axis represents the vector index, and the y-axis represents the vector values. Each color-coded line corresponds to a different value of $\alpha$ ranging from 0 to 1, illustrating how the projection vectors vary with $\alpha$. On the right are the spike rasters for two datasets. \textbf{b.} ReBaCCA values (purple curve) as a function of $\alpha$, with the x-axis representing $\alpha$ from 0 to 1. The blue and orange bars represent the variance explained by the first aligned dimension for Dataset 1 and Dataset 2.}
\label{Fig:different alpha results}
\end{figure}

We further investigate the influence of the continuum parameter $\alpha$ on the projection space and the ReBaCCA values. Figure \ref{Fig:different alpha results}\textbf{a} visualizes the optimal projection vectors $w_1^{(1)^*}$ and $w_2^{(1)^*}$ under different $\alpha$ values. The projection vectors emphasize distinct subsets of neurons with unique firing patterns. For example, in Case 1, the projection vector $w_1^{(1)^*}$ changes at index 10 while the projection vector $w_2^{(1)^*}$ changes at index 16. Similar phenomenon is also observed in Case 2: neurons with similar firing patterns are assigned comparable weights in the projection vectors. When $\alpha$ is restricted to certain intervals (e.g., from 0.6 to 1), the projection vectors remain consistent. When $\alpha$ varies across its full range, the resulting projection vectors are more diverse.

Despite the differences in the projection space, our proposed ReBaCCA values remain relatively stable across different values of  $\alpha$, as shown in Figure \ref{Fig:different alpha results}\textbf{b}. As noted earlier, $\alpha$ controls the trade-off between maximizing the correlation and preserving variance. When $\alpha=0$, the method focuses predominantly on maximizing correlation after the projection. Consequently, although the correlation is high, the variance explained is relatively low. As $\alpha$ increases, the correlation after alignment decreases, but the variance explained in the leading dimensions grows. Overall, these competing effects balance out, causing ReBaCCA to remain similar across a wide range of $\alpha$ values.

\section{Experimental Results}

\subsection{Delayed Nonmatch to Sample (DNMS) Task}

Having demonstrated the effectiveness of ReBaCCA-ss in accurately reconstructing latent dynamics from simulated data, we apply our method to real neural recordings collected during the Delayed Nonmatch to Sample (DNMS) task \citep{song2009nonlinear,song2014extraction}. All experimental procedures involving animals were reviewed and approved by the Institutional Animal Care and Use Committee at Wake Forest University, ensuring compliance with guidelines from the US Department of Agriculture, the International Association for the Assessment and Accreditation of Laboratory Animal Care, and the National Institutes of Health. 

\begin{figure}[!htbp]
\centering
\includegraphics[width=.9\textwidth]{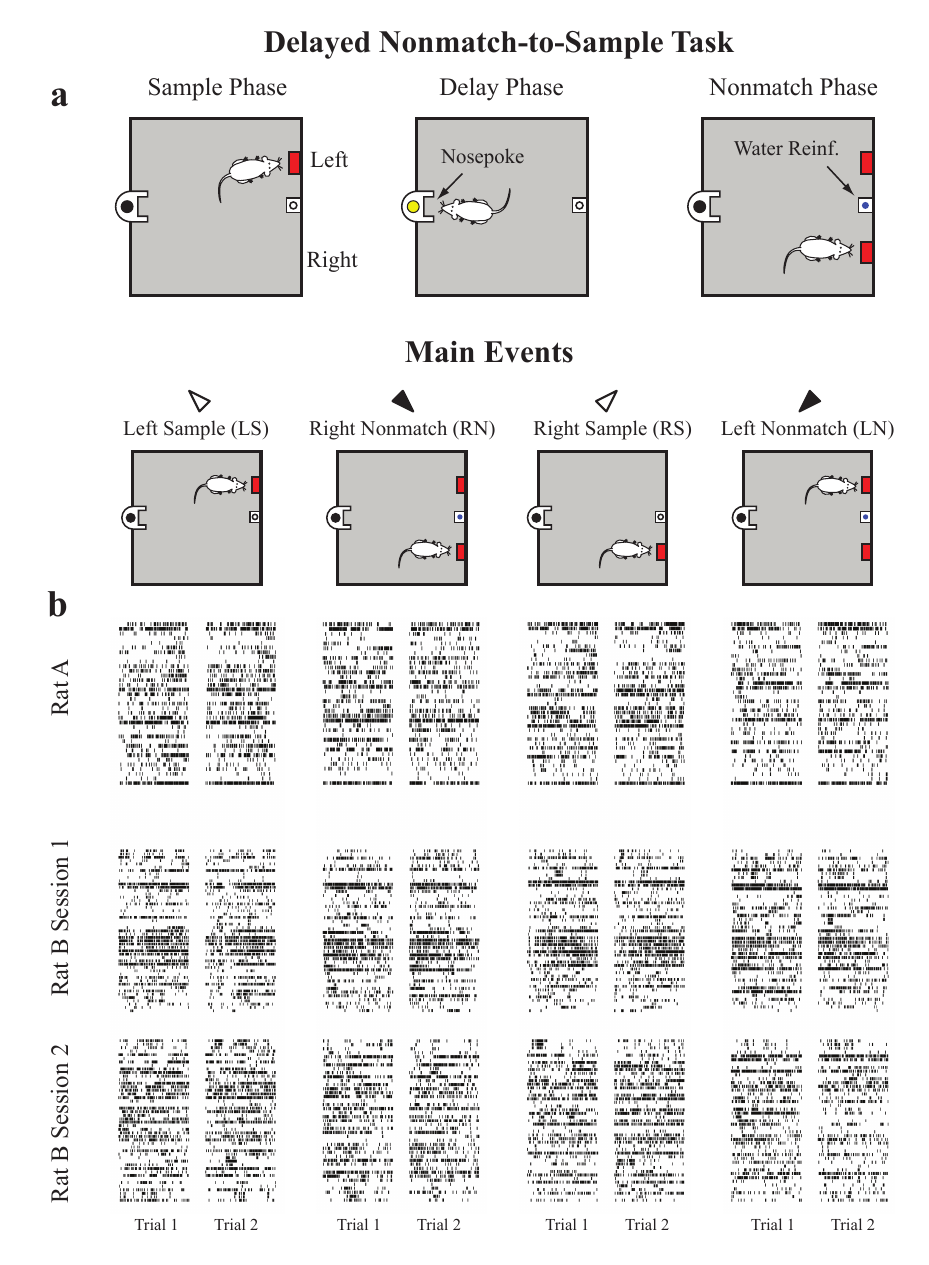}
\caption{Delayed Nonmatch to Sample Task (DNMS) data used in this study. \textbf{a.} Experimental paradigm of the DNMS task, illustrating the main behavioral events: Left Sample (LS), Right Nonmatch (RN), Right Sample (RS), and Left Nonmatch (LN). Each event is represented by a distinct triangular symbol and is defined by the spatial location (left vs. right) and task phase (sample vs. nonmatch). \textbf{b.} Spike raster plots illustrating neural activity associated with each of the four events, recorded from Rat A and Rat B (Sessions 1 and 2). For each event, spike patterns from two trials per session are shown, with the event time aligned to the center of the time axis.}
\label{Fig:dnms task and raster}
\end{figure}

In the DNMS task, male Long-Evans rats were trained to perform the task using a two-lever apparatus with randomized delay intervals, following protocols adapted from previous studies \citep{deadwyler1996hippocampal,hampson1999distribution}. During the sample phase of each trial, as shown in Figure \ref{Fig:dnms task and raster}\textbf{a}, the animal was trained to press one of the two levers presented in either the left or right position, referred to as the sample response. Following this, the lever was retracted, and a delay phase began. During this phase, the rat had to maintain a nose-poke in a lighted port located on the opposite wall, with the delay period varying randomly in duration. At the end of the delay phase, the nose-poke light was extinguished, both levers were extended again, and the rat had to press the lever opposite to the one pressed during the sample phase, known as the non-match response. Successful completion of the task resulted in a water reward. Each session consisted of approximately 100 trials, with each successful trial consisting of one of two pairs of behavioral events: left sample (LS) followed by right non-match (RN), or right sample (RS) followed by left non-match (LN).

Spike trains were recorded from the hippocampus using multi-electrode arrays (MEAs) while rats performed the DNMS task, which requires memory-dependent actions. Neural activity and corresponding behavioral events were recorded simultaneously during the task. Spikes were binned at 1 ms resolution and extracted within a ±2.5-second window around each behavioral event (LS, RN, RS, or LN). For analysis, we used three spike train datasets: one session from Rat A and two sessions from Rat B. Example spike raster plots aligned to the behavioral events, are shown in Figure \ref{Fig:dnms task and raster}\textbf{b}.

\begin{figure}[!htbp]
\centering
\includegraphics[width=\textwidth]{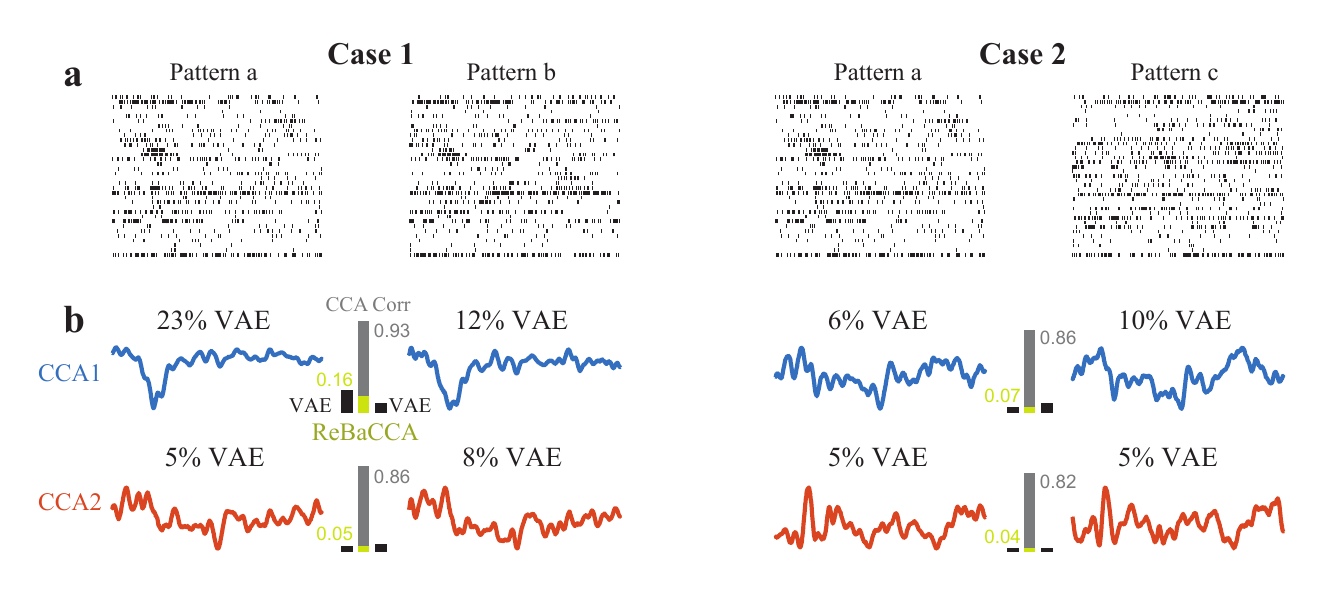}
\caption{Calculating ReBaCCA between spike patterns recorded during the DNMS task. \textbf{a.} Spike raster plots for three representative patterns from the same animal. Each plot shows spike times across multiple neurons, with rows corresponding to individual neurons and columns to time bins. Patterns a and b are from distinct trials of the RN event, while pattern c is from a separate trial of the LS event. \textbf{b.} Projection of the neural data onto the first two aligned dimensions following kernel smoothing. Blue and orange curves depict the temporal evolution of the projections for each pattern. The gray bars indicate canonical correlation (CCA) values, the black bars represent the proportion of variance explained by each component in the original spike matrices, and the yellow bars show the ReBaCCA values for each dimension.}
\label{Fig:DNMS CCA}
\end{figure}

ReBaCCA-ss is applied to compare spike patterns recorded during different DNMS trials (Figure \ref{Fig:DNMS CCA}\textbf{a}). In this example, in Case 1, Trials a and b exhibit similar firing patterns, whereas in Case 2, Trials a and c display relatively distinct patterns. However, when CCA is directly applied to align these spike patterns following kernel smoothing (Figure \ref{Fig:DNMS CCA}\textbf{b}), the correlation coefficients remain high for both cases (gray bars), without distinguishing the different levels of similarities in the two cases. This arises because CCA prioritizes maximizing correlation without considering the variance explained (black bars) in each aligned dimension. In contrast, our proposed ReBaCCA metric (yellow bar), which scales the correlation between the two patterns with the variance explained of each pattern, offers a more informative measure of similarity between spike patterns. In Case 1, the ReBaCCA value exceeds that of Case 2, providing a more robust evaluation of pattern similarity compared to the standalone CCA.

\subsection{Multidimensional Scaling (MDS) Visualization of the Pairwise Trial Similarity}

\begin{figure}[!htbp]
\centering
\includegraphics[width=.85\textwidth]{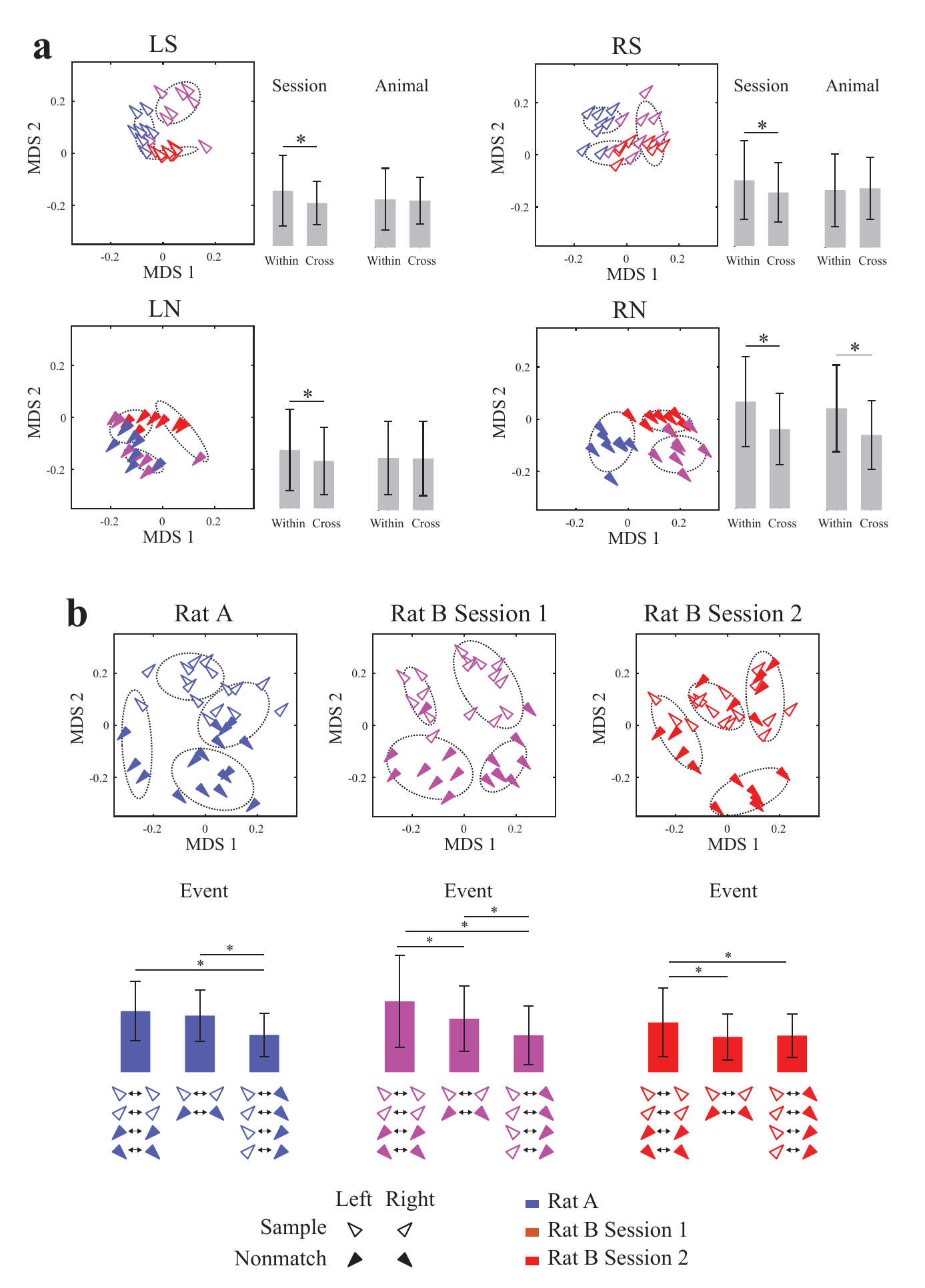}
\caption{Similarity between spike patterns across events, trials, sessions, and animals revealed by ReBaCCA-ss and MDS. Color represents sessions. Triangle type represents different events. \textbf{a.} MDS plot of patterns from the same event across different trials and sessions. Overlapping clusters indicate similar spike patterns. The bar plots present two comparisons: (1) within-session similarity versus cross-session similarity, and (2) within-animal similarity versus cross-animal similarity. \textbf{b.} MDS plot of patterns from the same session across different trials and events. The bar plots show three comparisons: (1) within-event (events with the same location and phase of the DNMS task) similarity, (2) cross-similar event (events with the same location or phase, e.g., LS vs. RS and LN vs. RN) similarity, and (3) cross-dissimilar event (events with different location and phase, e.g., LS vs. RN and RS vs. LN) similarity. For all bar plot comparisons, significant differences are indicated by t-test results ($p<0.05$, denoted by *).}
\label{Fig:MDS}
\end{figure}

The ReBaCCA framework is designed for pairwise comparisons between spike patterns. To obtain a comprehensive view of pattern similarity across conditions, ReBaCCA-ss values are computed for all possible pairs of spike patterns across trials, events, sessions, and animals. The resulting pairwise similarity matrix is then embedded into a low-dimensional space using Multidimensional Scaling (MDS) \citep{carroll1998multidimensional}, which positions each spike pattern as a point while preserving the pairwise ReBaCCA-ss relationships. This approach enables intuitive visualization of spike pattern similarities both within and across experimental conditions.

Three recording sessions from two animals were included in the analysis. For each of the four behavioral events, eight trials per session were randomly selected, yielding a total of 96 spike patterns. ReBaCCA-ss values were calculated for all possible pairs among these patterns, resulting in a 96 × 96 pairwise similarity matrix, where each element represents a ReBaCCA-ss value. MDS was then used to project this similarity matrix into a two-dimensional space for visualization (Figure \ref{Fig:MDS}).

Figure \ref{Fig:MDS}\textbf{a} shows that spike patterns corresponding to the same event tend to cluster by session, as indicated by the consistent colors within the dashed circle. This observation is supported by the accompanying bar plot, which demonstrates that within-session similarity is significantly higher than cross-session similarity. This result is consistent with the expectation that spike patterns recorded within the same session are more similar than those from different sessions. In contrast, when evaluating animal-wise similarity (red and purple vs. blue), within-animal similarity is not significantly different from cross-animal similarity, except for the RN event.

Figure \ref{Fig:MDS}\textbf{b} illustrates that, within a single session, spike patterns from different trials of the same event form tight clusters, as indicated by the uniform triangular markers within dashed circles. The bar plot confirms that within-event similarity significantly exceeds that of other comparisons. This finding supports the expectation that spike patterns evoked by the same event exhibit the highest degree of similarity across trials, reflecting shared neural features. Additionally, notable similarities are observed between certain event types—namely, between sample events (LS vs. RS) and between nonmatch events (LN vs. RN)—particularly in the data from Rat A and Rat B Session 1. These patterns suggest that these event pairs may share underlying neural representations or behavioral contexts. In contrast, spike patterns from unrelated event combinations exhibit the lowest similarity.

Together, these results demonstrate that ReBaCCA-ss, combined with MDS, provides an effective framework for quantifying and visualizing neural similarity across trials, events, sessions, and animals.
\newpage
\section{Discussion}
In this study, we introduce a novel framework for quantifying similarity between two population spike patterns, comprising three main steps (Figure \ref{Fig:algorithm flowchart}). First, ReBaCCA projects spike matrices into a latent space using CCC, generating a single similarity metric by weighting cross-dataset correlations according to the variance explained. Second, ReBaCCA-ss refines this metric by subtracting the similarity of smoothed surrogate spike trains from that of the original smoothed spike trains, thereby isolating the intrinsic similarity between spike patterns. Third, the optimal kernel bandwidth is selected by maximizing this difference.

A crucial parameter in our approach is $\alpha$, which controls the trade-off between maximizing cross-dataset correlation and variance explained in the spike patterns. $\alpha$ defines the projection directions that lie between those of CCA ($\alpha=0$) and PCA ($\alpha=1$) \citep{xie2020optimizing}, leading to distinct projection spaces (Figure \ref{Fig:different alpha results}\textbf{a}). Notably, the ReBaCCA values remain stable across different projection directions as shown in Figure \ref{Fig:different alpha results}\textbf{b}, demonstrating its robustness to variations in the projection space.  In this study, we use $\alpha=0.5$ to balance correlation alignment and variance explained in the final ReBaCCA-ss calculation. In practice, smaller values of $\alpha$ (e.g., $\alpha<0.33$) may be chosen to emphasize alignment across datasets, while larger values $\alpha$ (e.g., $\alpha>0.66$) may be preferable when prioritizing variance explained. Exploring a range of $\alpha$ values can provide a more comprehensive understanding of the relationships between two spike patterns and may reveal unique insights.

Surrogate matrices have been widely employed in neuroscience research to distinguish genuine neural patterns from artifacts. For example, in \citep{elsayed2017structure}, a framework was developed to assess whether observed population-level neural structures represent novel findings or merely reflect simpler, known features. In this study, it is demonstrated that CCA correlation increases following kernel smoothing, even for independent spike matrices (Figure \ref{Fig:cca theoretical corr}). Theoretical kernel-induced correlation resulting from CCA under specific conditions is also derived (see Appendix). In practical applications, randomly permuted spike matrices are used as a baseline to determine whether the observed similarity reflects true underlying structure or is an artifact introduced by the smoothing process.

Our proposed framework for pairwise similarity analysis between spike matrices provides a foundation for broader applications in neural data analysis. For instance, in Figure \ref{Fig:MDS}, we demonstrate how Multidimensional Scaling (MDS) can be used to visualize spike pattern similarities. Comparing spike pattern similarities within the same animal across different tasks may offer insights into neural adaptability and generalization across behavioral contexts. Similarly, comparing neural activity across animals performing the same task \citep{gallego2020long} can reveal shared neural dynamics that transcend individual variability. A systematic investigation of these applications is planned for future studies.

In addition, this method can be applied to validate large-scale models of biological systems \citep{hendrickson2015million,gene2019axonal}. By quantifying the similarity between real neural spike patterns and those generated by simulations, the accuracy of the models can be assessed \citep{williamson2019bridging}. This comparison may also yield novel error signals that can guide further optimization and refinement of the biological models.

One limitation of the current framework is the use of a single kernel bandwidth to smooth both spike matrices. As demonstrated, kernel bandwidth plays a critical role in capturing temporal features for cross-dataset comparison. When the temporal structures of the two datasets differ substantially, a single bandwidth may not be sufficient to represent both datasets accurately. In such cases, applying distinct kernel bandwidths to each dataset may be necessary for improved alignment and comparison.

Another limitation lies in the computational scalability of pairwise comparisons. Since the number of comparisons grows quadratically with the number of trials, the computational burden can become a bottleneck in large-scale experiments involving thousands of trials. To address this, future work may incorporate more efficient trial sampling techniques and leverage parallel computing to reduce computational demands.

\section*{Data and Code Availability}
The code, along with example data for demonstrating the modeling methodologies and reproducing figures in this manuscript, is publicly available at:
https://github.com/neural-modeling-and-interface-lab/ReBaCCA-ss

\section*{Acknowledgments}
This research was supported by NIH/NIDA BRAIN Initiative - Theories, Models and Methods (TMM) program (RF1DA055665/1R01EB031680).

\appendix
\section{Appendix}

\subsection{Deduction of the Average Correlation of Two Independent Spike Matrices After Smoothing and CCA}

For this deduction, we make several key assumptions to ensure the validity of our analytical results. First, we assume that the firing rate of each neuron is sufficiently low such that the average spike count per time bin is close to zero. Second, the kernel size used for smoothing cannot be excessively large relative to the total trial duration; otherwise, the smoothing would obscure temporal structure and reduce the effective number of independent observations. Finally, we assume that the number of neurons $N$ is not excessively large, so that the dimensionality of the spike matrices does not overwhelm the temporal relationship.

Denote two independent Poisson spike trains as
$x_1(t) = \sum_{k=1}^{M_1}\delta(t-t_k)$ and $x_2(t) = \sum_{m=1}^{M_2}\delta(t-t_m)$. $M_1$ and $M_2$ represent the number of spikes in $[0, T]$, $t_k$ and $t_m$ are spike times. The firing rates for two spike trains are denoted as $\lambda_1$ and $\lambda_2$, respectively. After kernel smoothing with the Gaussian kernel $g_\sigma(t)=\frac{1}{\sqrt{2\pi}\sigma}e^{-\frac{t^2}{2\sigma^2}}$, two spike trains are smoothed to: 

$s_1(t) = \int_0^T u_1(\tau)g_\sigma(t-\tau)d\tau = \sum_{k=1}^{M_1} g_\sigma(t-t_k)$ 

$s_2(t) = \int_0^T u_2(\tau)g_\sigma(t-\tau)d\tau = \sum_{m=1}^{M_2} g_\sigma(t-t_m)$

The variance of the smoothed spike trains can be calculated according to Campbell Theorem \citep{baddeley2007spatial}
\begin{align}
    \operatorname{var}(s_1(t)) &= E[s_1(t)^2] - E[s_1(t)]^2 \\
                &= E[\Big(\sum_{k=1}^{N_1} g_\sigma(t-t_k)\Big)^2] - \lambda_1^2 \\
                &= E[\sum_{k=1}^{N_1}g_\sigma(t-t_k)^2 + \sum_{k\neq m}g_\sigma(t-t_k)g_\sigma(t-t_m)] - \lambda_1^2 \\
                &= \lambda_1\int_0^Tg_\sigma(t-\tau)^2d\tau + \lambda_1\int_0^Tg_\sigma(t-\tau)d\tau\cdot\lambda_1\int_0^Tg_\sigma(t-\tau)d\tau - \lambda_1^2 \\
                &= \frac{\lambda_1}{2\sqrt{\pi}}\cdot\frac{1}{\sigma} \\
    \operatorname{var}(s_2(t)) &= \frac{\lambda_2}{2\sqrt{\pi}}\cdot\frac{1}{\sigma}
\end{align}

For Poisson spike trains with firing rate $\lambda$, the autocorrelation function is $R(\tau) = \lambda\delta(\tau) + \lambda^2$. After smoothing with the Gaussian kernel $g_\sigma(t)$, the autocorrelations become: 
\begin{align}
    R_{s_1}(\tau) &= \lambda_1^2 + \lambda_1\cdot e^{-\frac{\tau^2}{4\sigma^2}} \\
    R_{s_2}(\tau) &= \lambda_2^2 + \lambda_2\cdot e^{-\frac{\tau^2}{4\sigma^2}}
\end{align}

Since $x_1(t)$ and $x_2(t)$ are independent, the expected value of the sample correlation coefficient is 0. The variance of the sample correlation coefficient $\hat \rho$ is:
\begin{align}
    \operatorname{var}[\hat \rho] &= \frac{E[(\frac{1}{T}\int_0^Ts_1(t)s_2(t)dt)^2]}    {\operatorname{var}[s_1(t)]\cdot \operatorname{var}[s_2(t)]} \\
        &= \frac{E[\frac{1}{T^2}\int_0^T\int_0^T s_1(t)s_2(t)s_1(k)s_2(k) dtdk ] }
        {\operatorname{var}[s_1(t)]\cdot \operatorname{var}[s_2(t)]} \\
        &= \frac{\frac{1}{T^2}\int_0^T\int_0^T (R_1(t-k)-\lambda_1^2)(R_2(t-k)-\lambda_2^2) dtdk}{\operatorname{var}[s_1(t)]\cdot \operatorname{var}[s_2(t)]} \\
        &\approx \frac{\frac{1}{T}\int_0^T (R_1(\tau)-\lambda_1^2)(R_2(\tau)-\lambda_2^2) d\tau}{(\operatorname{var}[s_1(t)]\cdot \operatorname{var}[s_2(t)]} \\
        &=\sqrt{2\pi}\cdot \frac{\sigma}{T}
\end{align}

Thus, the distribution of the sample correlation coefficient $\hat\rho$ across many independent realizations is approximately Gaussian with mean 0 and variance $\sqrt{2\pi}\cdot \frac{\sigma}{T}$. When CCA is applied, all negative sample correlations become positive, therefore, the expected correlation coefficient after CCA $E[|\hat\rho|]$ is:
\begin{align}
    E[|\hat\rho|] &= \int_{-\infty}^\infty |x| \frac{1}{\sqrt{2\pi}\sigma}e^{-\frac{x^2}{2\sigma^2}}dx \\
        &=2\int_0^{\infty} x \frac{1}{\sqrt{2\pi}\sigma}e^{-\frac{x^2}{2\sigma^2}}dx \\
        &=\int_0^{\infty} \frac{1}{\sqrt{2\pi}\sigma}e^{-\frac{x^2}{2\sigma^2}}dx^2 \\
        &=(\frac{8}{\pi})^{1/4}\cdot\sqrt{\frac{\sigma}{T}}
        \label{eq:single spike train cc}
\end{align}

When generalizing to the case of two spike matrices, each with $N$ dimensions (i.e., $N$ neurons), the variance of the average sample correlation coefficient across all dimensions increases proportionally with $N$. After kernel smoothing, this variance is given by \citep{martin1979multivariate}
\begin{equation}
    \operatorname{var}_{\mathrm{mat}}[\hat{\rho}] = N \cdot \operatorname{var}[\hat{\rho}] = \sqrt{2\pi}\, \frac{N\sigma}{T}
\end{equation}
Therefore, following the same derivation as for the single spike train case (\ref{eq:single spike train cc}), the average correlation after smoothing and CCA for independent spike train matrices with $N$ neurons is:
\begin{equation}
    (\frac{8}{\pi})^{1/4}\cdot\sqrt{\frac{N\sigma}{T}}
\end{equation}

\singlespacing
\bibliography{reference}

\begin{thebibliography}{}

\bibitem[\protect\astroncite{Baddeley et~al.}{2007}]{baddeley2007spatial}
Baddeley, A., B{\'a}r{\'a}ny, I., and Schneider, R. (2007).
\newblock Spatial point processes and their applications.
\newblock {\em Stochastic Geometry: Lectures Given at the CIME Summer School Held in Martina Franca, Italy, September 13--18, 2004}, pages 1--75.

\bibitem[\protect\astroncite{Bj{\"o}rkstr{\"o}m and Sundberg}{1999}]{bjorkstrom1999generalized}
Bj{\"o}rkstr{\"o}m, A. and Sundberg, R. (1999).
\newblock A generalized view on continuum regression.
\newblock {\em Scandinavian Journal of Statistics}, 26(1):17--30.

\bibitem[\protect\astroncite{Carroll and Arabie}{1998}]{carroll1998multidimensional}
Carroll, J.~D. and Arabie, P. (1998).
\newblock Multidimensional scaling.
\newblock {\em Measurement, judgment and decision making}, pages 179--250.

\bibitem[\protect\astroncite{Chen and Fang}{2023}]{chen2023recent}
Chen, H. and Fang, Y. (2023).
\newblock Recent developments in implantable neural probe technologies.
\newblock {\em Mrs Bulletin}, 48(5):484--494.

\bibitem[\protect\astroncite{Chen et~al.}{2019}]{chen2019stability}
Chen, Y., Xin, Q., Ventura, V., and Kass, R.~E. (2019).
\newblock Stability of point process spiking neuron models.
\newblock {\em Journal of computational neuroscience}, 46:19--32.

\bibitem[\protect\astroncite{Deadwyler et~al.}{1996}]{deadwyler1996hippocampal}
Deadwyler, S.~A., Bunn, T., and Hampson, R.~E. (1996).
\newblock Hippocampal ensemble activity during spatial delayed-nonmatch-to-sample performance in rats.
\newblock {\em Journal of Neuroscience}, 16(1):354--372.

\bibitem[\protect\astroncite{Elsayed and Cunningham}{2017}]{elsayed2017structure}
Elsayed, G.~F. and Cunningham, J.~P. (2017).
\newblock Structure in neural population recordings: an expected byproduct of simpler phenomena?
\newblock {\em Nature neuroscience}, 20(9):1310--1318.

\bibitem[\protect\astroncite{Gallego et~al.}{2020}]{gallego2020long}
Gallego, J.~A., Perich, M.~G., Chowdhury, R.~H., Solla, S.~A., and Miller, L.~E. (2020).
\newblock Long-term stability of cortical population dynamics underlying consistent behavior.
\newblock {\em Nature neuroscience}, 23(2):260--270.

\bibitem[\protect\astroncite{Gene et~al.}{2019}]{gene2019axonal}
Gene, J.~Y., Bouteiller, J.-M.~C., Song, D., and Berger, T.~W. (2019).
\newblock Axonal anatomy optimizes spatial encoding in the rat entorhinal-dentate system: a computational study.
\newblock {\em IEEE Transactions on Biomedical Engineering}, 66(10):2728--2739.

\bibitem[\protect\astroncite{Hampson et~al.}{1999}]{hampson1999distribution}
Hampson, R.~E., Simeral, J.~D., and Deadwyler, S.~A. (1999).
\newblock Distribution of spatial and nonspatial information in dorsal hippocampus.
\newblock {\em Nature}, 402(6762):610--614.

\bibitem[\protect\astroncite{Hendrickson et~al.}{2015}]{hendrickson2015million}
Hendrickson, P.~J., Gene, J.~Y., Song, D., and Berger, T.~W. (2015).
\newblock A million-plus neuron model of the hippocampal dentate gyrus: critical role for topography in determining spatiotemporal network dynamics.
\newblock {\em IEEE Transactions on Biomedical Engineering}, 63(1):199--209.

\bibitem[\protect\astroncite{Huang et~al.}{2022}]{huang2022extracting}
Huang, Y., Zhang, X., Shen, X., Chen, S., Principe, J.~C., and Wang, Y. (2022).
\newblock Extracting synchronized neuronal activity from local field potentials based on a marked point process framework.
\newblock {\em Journal of Neural Engineering}, 19(4):046043.

\bibitem[\protect\astroncite{Jonsson et~al.}{2016}]{jonsson2016bioelectronic}
Jonsson, A., Inal, S., Uguz, I., Williamson, A.~J., Kergoat, L., Rivnay, J., Khodagholy, D., Berggren, M., Bernard, C., Malliaras, G.~G., et~al. (2016).
\newblock Bioelectronic neural pixel: Chemical stimulation and electrical sensing at the same site.
\newblock {\em Proceedings of the National Academy of Sciences}, 113(34):9440--9445.

\bibitem[\protect\astroncite{Lee}{2007}]{lee2007continuum}
Lee, M.~H. (2007).
\newblock {\em Continuum direction vectors in high dimensional low sample size data}.
\newblock The University of North Carolina at Chapel Hill.

\bibitem[\protect\astroncite{Marek et~al.}{2022}]{marek2022reproducible}
Marek, S., Tervo-Clemmens, B., Calabro, F.~J., Montez, D.~F., Kay, B.~P., Hatoum, A.~S., Donohue, M.~R., Foran, W., Miller, R.~L., Hendrickson, T.~J., et~al. (2022).
\newblock Reproducible brain-wide association studies require thousands of individuals.
\newblock {\em Nature}, 603(7902):654--660.

\bibitem[\protect\astroncite{Martin and Maes}{1979}]{martin1979multivariate}
Martin, N. and Maes, H. (1979).
\newblock Multivariate analysis.
\newblock {\em London, UK: Academic}.

\bibitem[\protect\astroncite{Pillow and Aoi}{2017}]{pillow2017population}
Pillow, J.~W. and Aoi, M.~C. (2017).
\newblock Is population activity more than the sum of its parts?
\newblock {\em Nature neuroscience}, 20(9):1196--1198.

\bibitem[\protect\astroncite{Safaie et~al.}{2023}]{safaie2023preserved}
Safaie, M., Chang, J.~C., Park, J., Miller, L.~E., Dudman, J.~T., Perich, M.~G., and Gallego, J.~A. (2023).
\newblock Preserved neural dynamics across animals performing similar behaviour.
\newblock {\em Nature}, 623(7988):765--771.

\bibitem[\protect\astroncite{Saxena and Cunningham}{2019}]{saxena2019towards}
Saxena, S. and Cunningham, J.~P. (2019).
\newblock Towards the neural population doctrine.
\newblock {\em Current opinion in neurobiology}, 55:103--111.

\bibitem[\protect\astroncite{Scholten et~al.}{2023}]{scholten2023shared}
Scholten, K., Xu, H., Song, D., and Meng, E. (2023).
\newblock A shared resource for building polymer-based microelectrode arrays as neural interfaces.
\newblock In {\em 2023 11th International IEEE/EMBS Conference on Neural Engineering (NER)}, pages 1--4. IEEE.

\bibitem[\protect\astroncite{She et~al.}{2022}]{she2022double}
She, X., Berger, T.~W., and Song, D. (2022).
\newblock A double-layer multi-resolution classification model for decoding spatiotemporal patterns of spikes with small sample size.
\newblock {\em Neural computation}, 34(1):219--254.

\bibitem[\protect\astroncite{She et~al.}{2024}]{she2024distributed}
She, X., Moore, B.~J., Roeder, B.~M., Nune, G., Robinson, B.~S., Lee, B., Shaw, S., Gong, H., Heck, C.~N., Popli, G., et~al. (2024).
\newblock Distributed temporal coding of visual memory categories in human hippocampal neurons.
\newblock {\em Research Square}, pages rs--3.

\bibitem[\protect\astroncite{Song et~al.}{2009}]{song2009nonlinear}
Song, D., Chan, R.~H., Marmarelis, V.~Z., Hampson, R.~E., Deadwyler, S.~A., and Berger, T.~W. (2009).
\newblock Nonlinear modeling of neural population dynamics for hippocampal prostheses.
\newblock {\em Neural Networks}, 22(9):1340--1351.

\bibitem[\protect\astroncite{Song et~al.}{2014}]{song2014extraction}
Song, D., Harway, M., Marmarelis, V.~Z., Hampson, R.~E., Deadwyler, S.~A., and Berger, T.~W. (2014).
\newblock Extraction and restoration of hippocampal spatial memories with non-linear dynamical modeling.
\newblock {\em Frontiers in Systems Neuroscience}, 8:97.

\bibitem[\protect\astroncite{Stone and Brooks}{1990}]{stone1990continuum}
Stone, M. and Brooks, R.~J. (1990).
\newblock Continuum regression: cross-validated sequentially constructed prediction embracing ordinary least squares, partial least squares and principal components regression.
\newblock {\em Journal of the Royal Statistical Society: Series B (Methodological)}, 52(2):237--258.

\bibitem[\protect\astroncite{Sundberg}{1993}]{sundberg1993continuum}
Sundberg, R. (1993).
\newblock Continuum regression and ridge regression.
\newblock {\em Journal of the Royal Statistical Society: Series B (Methodological)}, 55(3):653--659.

\bibitem[\protect\astroncite{Truccolo et~al.}{2005}]{truccolo2005point}
Truccolo, W., Eden, U.~T., Fellows, M.~R., Donoghue, J.~P., and Brown, E.~N. (2005).
\newblock A point process framework for relating neural spiking activity to spiking history, neural ensemble, and extrinsic covariate effects.
\newblock {\em Journal of neurophysiology}, 93(2):1074--1089.

\bibitem[\protect\astroncite{Williamson et~al.}{2019}]{williamson2019bridging}
Williamson, R.~C., Doiron, B., Smith, M.~A., and Yu, B.~M. (2019).
\newblock Bridging large-scale neuronal recordings and large-scale network models using dimensionality reduction.
\newblock {\em Current opinion in neurobiology}, 55:40--47.

\bibitem[\protect\astroncite{Xie et~al.}{2020}]{xie2020optimizing}
Xie, Z., Chen, X., Huang, G., et~al. (2020).
\newblock Optimizing a vector of shrinkage factors for continuum regression.
\newblock {\em Chemometrics and Intelligent Laboratory Systems}, 206:104141.

\bibitem[\protect\astroncite{Xu et~al.}{2018}]{xu2018acute}
Xu, H., Hirschberg, A.~W., Scholten, K., Berger, T.~W., Song, D., and Meng, E. (2018).
\newblock Acute in vivo testing of a conformal polymer microelectrode array for multi-region hippocampal recordings.
\newblock {\em Journal of neural engineering}, 15(1):016017.

\bibitem[\protect\astroncite{Zhang et~al.}{2019}]{zhang2019clustering}
Zhang, X., Libedinsky, C., So, R., Principe, J.~C., and Wang, Y. (2019).
\newblock Clustering neural patterns in kernel reinforcement learning assists fast brain control in brain-machine interfaces.
\newblock {\em IEEE Transactions on Neural Systems and Rehabilitation Engineering}, 27(9):1684--1694.

\end{thebibliography}

\end{document}